\documentclass[12pt,a4paper]{article}
\usepackage[T1]{fontenc}
\usepackage[latin9]{inputenc}
\usepackage{float}
\usepackage{amsmath}
\usepackage{graphicx}
\usepackage{amssymb}

\usepackage{epsfig} 

\newcommand\plb[3]   {{\it Phys.\ Lett.\ }{\bf B #1} (#2) #3}
\newcommand\prd[3]   { {{\it Phys.\ Rev.\ }{\bf D #1} (#2) #3}} 
\newcommand\jhep[3]  {	{{\it J. High Energy Phys.\ }{\bf #1} (#2) #3}}
\newcommand\npb[3]   {{\it Nucl.\ Phys.\ }{\bf B #1} (#2) #3}

\newcommand\epjc[3]  {{\it Eur.\ Phys.\ J. }{\bf C #1} (#2) #3}
\newcommand\jphg[3]  {{\it J. Phys.\ }{\bf G #1} (#2) #3}

\begin{document}

\makeatother

\title{\textbf{\Large S and T Parameters in the Fermion Condensate Model
}}

\author{G. Cynolter$^{*}$, E. Lendvai$^{*}$  and
\begin{tabular}{|c|}
\hline
G. P\'ocsik$^{\dagger}$ \tabularnewline
\hline
\end{tabular}  
 }

\date{$^{*}$ Theoretical Physics Research Group of Hungarian Academy of
Sciences, E\"otv\"os University, Budapest, 1117 P\'azm\'any P\'eter s\'et\'any 1/A,
Hungary  \\
$^{\dagger}$ Institute for Theoretical Physics, E\"otv\"os University, Budapest, 1117 P\'azm\'any P\'eter s\'et\'any 1/A,
Hungary
}

\maketitle

\begin{abstract}
We calculate the oblique electroweak corrections and confront them with
the experiments in a composite Higgs version of the standard model.
A vector-like weak doublet and a singlet fermion are added to the standard 
model without elementary Higgs. 
Due to quartic coupling there is a mixing between the components
of the new fields triggering electroweak symmetry breaking. 
The Peskin-Takeuchi  ${S}$ and ${T}$ electroweak parameters are
presented.
The new sector of vector-like fermions is slightly constrained, 
${T}$ gives an upper bound on the mixing angle of the new fermions, which
is already constrained by self-consistent gap-equations.
${S}$ gives  no constraints on the masses.
This extension can give a positive contribution to ${T} $, allowing for a heavy Higgs boson in electroweak precision tests of the Standard Model. 
\end{abstract}


\section{Introduction}

There are strong indications and expectations that the LHC will reveal the physics of electroweak symmetry breaking. The Electroweak Precision Tests (EWPT) in the standard model favour a light Higgs 
($ m_H $ below appr. 200 GeV), but the UV sensitivity of the Higgs mass motivates the study of alternative models. 
The original technicolor idea \cite{Weinberg, Susskind} of fermion condensation is already thirty years old, but it still gives motivation for new research, see a recent review \cite{Hillpr}, Chivukula et al. in \cite{pdg} and references therein.
To provide  fermion masses extended technicolor gauge interactions (ETC) \cite{etc,etc2} must be included. The tension between sizeable quark masses and avoiding flavour changing neutral currents led to introduce  walking, near conformal dynamics \cite{holdom, bando}. These ideas and the phase diagram of strongly interacting models triggered activity in lattice studies \cite{latticeconf}, and further new technicolor models were constructed based on adjoint  or two index symmetric representations of the new fermions \cite{sannino}.
Inspired by discretized higher dimensional theories "Little Higgs" \cite{lH} models provide a new class of composite Higgs models. Higgsless models \cite{csaki} do not utilize a scalar Higgs boson, but using the AdS/CFT correspondence these are  extra dimensional "duals" of walking technicolor theories.

Viable models of Electroweak Symmetry Breaking must fulfill the EWPT.
The original technicolor theories with QCD-like dynamics gave large contribution
to the $S $ parameter \cite{stu}. New  technicolor theories can also provide very low S parameter in case of one chiral techni-fermion and walking, near conformal dynamics.
One can overcome the difficulties with the S parameter using fermions in vector representation.
In a recently proposed dynamical symmetry breaking model  vector-like fermions of different  $SU_L(2)$ representation (singlet and doublet) mix due to condensation \cite{fcm}. Mixing is essential in the model. Vector-like extension of the standard model is widely studied in the literature. They naturally appear in extra dimensional models with bulk
fermions e.g \cite{UED}, in little Higgs theories \cite{lH},
 in models of so called improved naturalness consistent with a heavy
 Higgs scalar \cite{improved} and in simple fermionic models of dark
 matter \cite{darkmatter,dm2,dm3}.
There are known results for precision electroweak parameters  for extra vector-like quarks 
\cite{silva,dawson,dmhiggs,fcmew}, but here the mixing of the fermions is a new phenomenon.
In this paper we calculate the Peskin-Takeuchi S and T parameters in the recently proposed fermion condensate  model based on vector-like fermions, taking into account the non-trivial mixing and the solution of the gap equation with a cutoff.

\section{Fermion Condensate Model}

In a recent paper \cite{fcm} self-interacting vector-like fermions were introduced
in the standard model instead of an elementary standard scalar Higgs.
The new colourless fermions  are an extra neutral weak $SU(2)$ singlet
$\Psi_{S}$ ($T=Y=0$) and a doublet $\Psi_{D}=\left(\begin{array}{c}
\Psi_{D}^{+}\\ \Psi_{D}^{0}\end{array}\right)$ with hypercharge 1.
New fermions like these  are often dubbed leptons, because they do not participate in
strong interactions. 
A model with similar fermion content were studied by Maekawa \cite{maek}.
There is a new $Z_{2}$ symmetry acting only on the new fermions, which protects them 
from mixings with  the standard model quarks and leptons.
The lightest new fermion is stable therefore it is an ideal weakly 
interacting dark matter candidate. 

The new fermions have the following kinetic terms, Dirac mass terms and quartic self-interactions
\begin{eqnarray}
L_{\Psi} & = & \phantom+i\overline{\Psi}_{D}D_{\mu}\gamma^{\mu}\Psi_{D}+i\overline{\Psi}_{S}\partial_{\mu}\gamma^{\mu}\Psi_{S}-m_{0D}\overline{\Psi}_{D}\Psi_{D}-m_{0S}\overline{\Psi}_{S}\Psi_{S}+\nonumber \\
 &  & 
+\lambda_{1}\left(\overline{\Psi}_{D}\Psi_{D}\right)^{2}+\lambda_{2}\left(\overline{\Psi}_{S}\Psi_{S}\right)^{2}+2\lambda_{3}\left(\overline{\Psi}_{D}\Psi_{D}\right)\left(\overline{\Psi}_{S}\Psi_{S}\right) .
\label{eq:4fermion}
\end{eqnarray}
$D_{\mu}$ is the covariant derivative
\begin{equation}
D_{\mu}=\partial_{\mu}-i\frac{g}{2}\underline{\tau}\,\underline{W}_{\mu}-i\frac{g'}{2}B_{\mu},\label{eq:covariantd}\end{equation}
where $\underline{W}_{\mu,}B_{\mu}$ and $g,\; g'$ are the standard
weak gauge boson fields and couplings, respectively. 
Equation (\ref{eq:4fermion}) describes non-renormalizable effective interactions. It is a low energy model valid up to a cutoff  $\Lambda\simeq 4\pi v \simeq 3$ TeV.
It was shown in ref. \cite{fcmgap} that if the $\lambda_3$ quartic coupling exceeds a critical value then  the four-fermion interactions in (\ref{eq:4fermion}) generate bilinear fermion condensates
\begin{eqnarray}
& & \left\langle \overline{\Psi}_{D\alpha}^{0}\Psi_{D\beta}^{0}\right\rangle _{0} =
 a_{1}\delta_{\alpha\beta},\label{eq:condD}\\
& & \left\langle \overline{\Psi}_{D\alpha}^{+}\Psi_{D\beta}^{+}\right\rangle _{0} =
 a_{+}\delta_{\alpha\beta},\label{eq:condP}\\
& & \left\langle \overline{\Psi}_{S\alpha}\Psi_{S\beta}\right\rangle _{0} = 
 a_{2}\delta_{\alpha\beta},\label{eq:condS}\\
& & \left\langle \overline{\Psi}_{S}\Psi_{D}\right\rangle _{0}=\left\langle \left(\begin{array}{c} \overline{\Psi}_{S}\Psi_{D}^{+} \\
 \overline{\Psi}_{S}\Psi_{D}^{0}\end{array}\right)\right\rangle _{0} \neq  
 0 .
 \label{eq:conddoublet}
\end{eqnarray}
The non-diagonal condensate in (\ref{eq:conddoublet})
spontaneously breaks the  $SU_{L}(2)\times U_{Y}(1)$ electroweak symmetry to $U_{em}(1)$. 
With the gauge transformations of $\Psi_{D}$ the condensate (\ref{eq:conddoublet}) can always be transformed into
a real lower component,
\begin{equation}
\left\langle \overline{\Psi}_{S\alpha}\Psi_{D\beta}^{0}\right\rangle _{0}=a_{3}\delta_{\alpha\beta},\quad\left\langle \overline{\Psi}{}_{S\alpha}\Psi_{D\beta}^{+}\right\rangle _{0}=0,
\label{eq:mixed cond}
\end{equation}
where $a_{3}$ is real.
The composite operator $\overline{\Psi}_{S}\Psi_{D}$ resembles the standard scalar doublet. 

The mixed condensate of $\overline{\Psi}_{S}\Psi_{D} $  generates masses for the  the standard fermions via the 
following four-fermion interactions:
\begin{equation}
L_{f} = 
 g_{f}\left(\overline{\Psi}_{L}^{f}\Psi_{R}^{f}\right)\left(\overline{\Psi}_{S}\Psi_{D}\right)
+g_{f}
\left(\overline{\Psi}_{R}^{f}\Psi_{L}^{f}\right)\left(\overline{\Psi}_{D}\Psi_{S}\right).
\label{eq:Yukawa}
\end{equation}
The neutrinos so  far massless; they can get masses introducing  right handed neutrinos, similarly as in the original standard model.
Here $ L_f$ generates masses 
for the leptons (f=$e, \mu, \tau $) in the linearized, or mean-field, approximation
\begin{equation}
m_f = -4g_f a_3.
\end{equation}
The weak gauge bosons receive their masses from the effective low energy interactions of 
an auxiliary composite scalar $\Phi=\overline{\Psi}_{S}\Psi_{D}$ \cite{fcm}. 
The new symmetry breaking sector possesses a global O(4) symmetry.
After electroweak symmetry breaking there is a residual O(3)$ \simeq $ SU(2) symmetry softly broken
by the electromagnetic interactions and the mass difference of the neutral and charged fermions
 \cite{fcmgap}. This custodial SU(2) ensures that 
$\rho_{\hbox{tree}} =1$ and it receives small corrections at one-loop level.

The dynamical condensates (\ref{eq:condD}-\ref{eq:condS}) contribute to the mass terms of the new fermions
in the Lagrangian \eqref{eq:4fermion}. The mixed condensate \eqref{eq:mixed cond} generates mixing between the new
fermions in the linearized approximation:
\begin{equation}
L_{\psi}\rightarrow -m_{+}\overline{\Psi_{D}^{+}}\Psi_{D}^{+}-m_{1}\overline{\Psi_{D}^{0}}\Psi_{D}^{0}-m_{2}\overline{\Psi}_{S}\Psi_{S}-m_{3}\left(\overline{\Psi^{0}}_{D}\Psi_{S}+\overline{\Psi}_{S}\Psi_{D}^{0}\right),\label{eq:fermion mass}
\end{equation}
where
\begin{eqnarray}
m_{+} & = & m_{0D}-6\lambda_{1}a_{+}-8\left(\lambda_{1}a_{1}+\lambda_{3}a_{2}\right)=m_{1}+2\lambda_{1}\left(a_{+}-a_{1}\right)\label{eq:mtablp}\\
m_{1} & = & m_{0D}-6\lambda_{1}a_{1}-8\left(\lambda_{1}a_{+}+\lambda_{3}a_{2}\right),\label{eq:mtable}\\
m_{2} & = & m_{0S}-6\lambda_{2}a_{2}-8\lambda_{3}\left(a_{1}+a_{+}\right),\label{eq:mtabl2}\\
m_{3} & = & 2\lambda_{3}a_{3}.\label{eq:mtabl3}
\end{eqnarray}

If $m_{3}$ does not vanish
(\ref{eq:fermion mass}) is diagonalized via unitary transformation
to get physical mass eigenstates 
\begin{eqnarray}
\Psi_{1} & = & \phantom{-}c\,\Psi_{D}^{0}+s\,\Psi_{S},\nonumber \\
\Psi_{2} & = & -s\,\Psi_{D}^{0}+c\,\Psi_{S},\label{eq:fermion mixing}
\end{eqnarray}
where $c=\cos\phi$ and $s=\sin\phi$, $\phi$ is the mixing angle.
The masses of the physical fermions $\Psi_{1},\:\Psi_{2}$ are
\begin{equation}
2M_{1,2}=m_{1}+m_{2}\pm\frac{m_{1}-m_{2}}{\cos2\phi}.\label{eq:mphys}
\end{equation}
The mixing angle is defined by 
\begin{equation}
2m_{3}=(m_{1}-m_{2})\tan2\phi.\label{eq:def phi}
\end{equation}
The original masses in terms of the physical masses are $m_{1}=  c^{2}M_{1}+s^{2}M_{2}$
and $m_{2}=  s^{2}M_{1}+c^{2}M_{2} $

The physical eigenstates themselves form condensates, but the diagonalization 
eliminated the mixed one:
\begin{eqnarray}
c^{2}\left\langle \overline{\Psi}_{1\alpha}\Psi_{1\beta}\right\rangle _{0}+s^{2}\left\langle \overline{\Psi}_{2\alpha}\Psi_{2\beta}\right\rangle _{0} & = & a_{1}\delta_{\alpha\beta},\\
s^{2}\left\langle \overline{\Psi}_{1\alpha}\Psi_{1\beta}\right\rangle _{0}+c^{2}\left\langle \overline{\Psi}_{2\alpha}\Psi_{2\beta}\right\rangle _{0} & = & a_{2}\delta_{\alpha\beta},\label{eq:condphys}\\
cs\left\langle \overline{\Psi}_{1\alpha}\Psi_{1\beta}\right\rangle _{0}-cs\left\langle \overline{\Psi}_{2\alpha}\Psi_{2\beta}\right\rangle _{0} & = & a_{3}\delta_{\alpha\beta}.
\end{eqnarray}

The equations (\ref{eq:mtablp}-\ref{eq:mtabl3}) can be formulated
as gap equations \cite{fcmgap} in terms of the physical fields expressing
both the masses and the condensates with $\Psi_{1}$, $\Psi_{2}$
and $\Psi_{+}\equiv\Psi_{D}^{+}$. Assuming vanishing original lagrangian masses,
$m_{0S}=0$, $m_{0D}=0$, the complete set of gap equations are
\begin{eqnarray}
c\cdot s\left(M_{1}-M_{2}\right) & = & 2\lambda_{3}\; c\cdot s\left(I_{1}-I_{2}\right),\label{eq:gap3}\\
c^{2}M_{1}+s^{2}M_{2} & = & -\lambda_{1}\left(6\left(c^{2}I_{1}+s^{2}I_{2}\right)+8I_{+}\right)-8\lambda_{3}\left(s^{2}I_{1}+c^{2}I_{2}\right),\label{eq:gap1}\\
s^{2}M_{1}+c^{2}M_{2} & = & -6\lambda_{2}\left(s^{2}I_{1}+c^{2}I_{2}\right)-8\lambda_{3}\left(c^{2}I_{1}+s^{2}I_{2}+I_{+}\right),\label{eq:gap2}\\
M_{+} & = & -\lambda_{1}\left(8\left(c^{2}I_{1}+s^{2}I_{2}\right)+6I_{+}\right)-8\lambda_{3}\left(s^{2}I_{1}+c^{2}I_{2}\right).
\label{eq:gap+}
\end{eqnarray}
Where $ I_i$ (i=1,2,+ ) are defined from the condensates. Approximating them by free field propagators
 \begin{equation}
\left\langle \overline{\Psi}_{i\alpha}\Psi_{i\beta}\right\rangle =\frac{\delta_{\alpha\beta}}{4}I_{i}=-\frac{\delta_{\alpha\beta}}{8\pi^{2}}M_{i}\left(\Lambda^{2}-M_{i}^{2}\ln\left(1+\frac{\Lambda^{2}}{M_{i}^{2}}\right)\right),\quad i=1,2,+,\label{eq:free1}
\end{equation}
where $M_{+}=m_{+}$. Here $\Lambda$ is a four-dimensional physical
cutoff, it sets the scale of the new physics responsible for the non-renormalizable
operators. $\Lambda$ is expected to be a loop factor higher than the scale of weak interactions, whose quanta ($ W^\pm, Z$) get their masses from the model,  $\Lambda \simeq 4\pi v \simeq 3$ TeV.

There are four equation (\ref{eq:gap3}-\ref{eq:gap+}) and four parameters, $M_1,\,  M_2,\, M_+,\, \cos \phi$. 
Equation (\ref{eq:gap3}) has  the form of a usual gap equation in terms of the mass difference $M_2-M_1$. 
The gap equations always have a symmetric solution with vanishing masses or mass difference in (\ref{eq:gap3}),
as generally $I_i \sim M_i$.
If $\lambda_3$ is negative and if  $|\lambda_3|$
exceeds a critical value,  $\pi^2/\Lambda^2 $ then there is a further, energetically 
favoured \cite{klev} symmetry breaking  solution ($M_{1}\neq M_{2}$).
In this case the non-diagonal $a_3 $ condensate  is formed, which triggers mixing between different representations of the weak gauge group; the electroweak symmetry is broken dynamically.
For $ \lambda_3 < -\pi^2/\Lambda^2 $ and $\lambda_3$  close to it's critical value the mass difference $|M_1-M_2|$ is much smaller than the cutoff $\Lambda$.
For $\lambda_3$ above the critical value $(M_1 - M_2) c\cdot s =0$.
In this case the physically relevant solution is  $c \cdot s=0$, there is no meaningful mixing, 
the electroweak symmetry is not broken.

For physical values of the mixing angle $0\leq c^{2}\leq1$ we get from the equations (\ref{eq:gap3}-\ref{eq:gap+})
\begin{equation}
M_{1}\leq M_{+}\leq M_{2}.\label{eq:mpconstr}
\end{equation}
There is also a critical value for $\lambda_{1,2}$.
Considering the limit $M_{+}\rightarrow M_{2}=M$ and $M_{1}\rightarrow 0$ we find 
for the massive solution
\begin{equation}
\lambda_{1}=\frac{1}{7}\frac{\pi^{2}}{\Lambda^{2}-M^{2}\ln\left(1+\frac{\Lambda^{2}}{M^{2}}\right)},\quad\lambda_{2}=\frac{4}{3}\frac{\pi^{2}}{\Lambda^{2}-M^{2}\ln\left(1+\frac{\Lambda^{2}}{M^{2}}\right)}.\label{eq:l1crit}
\end{equation}
Equation (\ref{eq:l1crit}) provides massive solutions
if $\lambda_{1}\geq\frac{1}{7}\frac{\pi^{2}}{\Lambda^{2}}$ and $\lambda_{2}\geq\frac{4}{3}\frac{\pi^{2}}{\Lambda^{2}}$.
It is remarkable  that the small mass solutions are found  not 
in the neighbourhood of the critical values, but for $\lambda_{1}\sim\frac{5}{7}\frac{\pi^{2}}{\Lambda^{2}}$
and $\lambda_{2}\sim3\frac{\pi^{2}}{\Lambda^{2}}$.
To get small mass difference $\lambda_3$ must be relatively close to it's critical value.

As the four-fermion interactions are non-renormalizable the values of the coupling constants are
constrained by perturbative unitarity, too \cite{unit,unit2}.
Consider the amplitudes of two particle 
elastic scattering processes (of the new fermions) and impose $\left|\Re a_{0}\right|\leq1/2$
for the $J=0$ partial wave amplitudes. The contact graph gives the
dominant contribution, neglecting the fermion masses for the $\Psi_{D}^{(+)}\Psi_{D}^{(-)}$
scattering gives an upper bound on $\lambda_{1}$ coupling.
The detailed analysis gives the same upper bound \cite{fcmgap}
\begin{equation}
\left| \lambda_{i} \right|s\leq8\pi \,,\quad i=1,\, 2,\, 3,
\end{equation}
where $s$ is the maximal center of mass energy of a given process majored by the 
general cutoff $\Lambda$.
\begin{figure}[t]
\begin{center}
\includegraphics[scale=0.50]{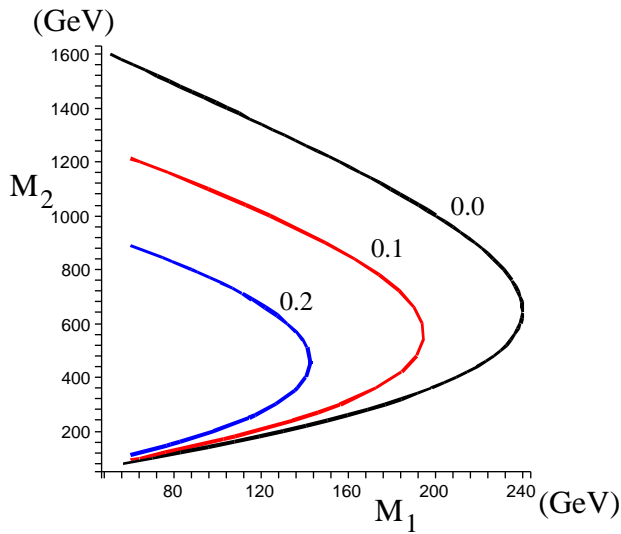}
\end{center}
\begin{center}
\caption{The maximum value of the cosine$^2$ of the mixing angle on the $M_1, \, M_2$ plane from the gap equation and unitarity. $c^2$ can be higher inside the curves.}
\end{center}
\end{figure}

The numerical solutions of the gap equations taking into account perturbative unitarity
give a window for the masses. The solution with degenerate masses ($M_1=M_+=M_2 $) goes with
$\lambda_2$ above the unitarity bound, it is not allowed. 
Generally the masses  constrained most severely by the unitarity of the coupling $\lambda_2$.
For $\Lambda=3$ TeV the lighter neutral fermion (we can choose it to be $\Psi_1$) 
must be fairly light $M_1 < 240 $ GeV.
The allowed  $M_1, \, M_2$ masses and  the maximum value of $c^2$ is shown in Figure 1.
The charged fermion mass must be relatively close to the mass of the heavier neutral one.
The mixing angle $\phi$ is relatively close to $\cos \phi \sim 0$, the mixing is weak, see the curve on the right in Figure 2.
$\Psi_{2}$ is mostly composed of $\Psi_{D}^{0}$ and there is only a small mass splitting in the
doublet $\Psi_{D}$ after symmetry breaking.

The collider phenomenology and radiative corrections in the model
are coming from the doublet kinetic term in (\ref{eq:4fermion}) taking
into account the mixing (\ref{eq:fermion mixing})

\begin{eqnarray}
L^{I} & = & \phantom{+}\overline{\Psi_{D}^{+}}\gamma^{\mu}\Psi_{D}^{+}\left(\frac{g'}{2}B_{\mu}+\frac{g}{2}W_{3\mu}\right)+\nonumber \\
 &  &
 +\left(c^{2}\overline{\Psi}_{1}\gamma^{\mu}\Psi_{1}+s^{2}\overline{\Psi}_{2}\gamma^{\mu}\Psi_{2}-sc 
\left(\overline{\Psi}_{1}\gamma^{\mu}\Psi_{2}+\overline{\Psi}_{2}\gamma^{\mu}\Psi_{1}\right)\right)\left(\frac{g'}{2}B_{\mu}-\frac{g}{2}W_{3\mu}\right)+\nonumber \\
 &  &
 +\left[\frac{g}{\sqrt{2}}W_{\mu}^{+}\left(c\overline{\Psi_{D}^{+}}\gamma^{\mu}\Psi_{1}-s
\overline{\Psi_{D}^{+}}\gamma^{\mu}\Psi_{2}\right)+h.c.\right].\label{eq:Llmix}
\end{eqnarray}
We will explore the consequences of these interactions in the decay of the $Z$ boson and 
the precision electroweak test of the standard model.

\subsection{New fermions constrained from the Z decay}

The proposed new fermions could not be seen in the high energy experiments so far,
because of their large masses and/or small couplings to ordinary particles.
The mixing in the doublet reduces the coupling to the gauge bosons, but the new charged fermion
is not affected. From the LEP1 and LEP2 measurements there is lower bound for
the mass of a heavy charged lepton, valid here $M_+ > 100$ GeV \cite{pdg}.
For the neutral component of the doublet (without mixing) there are smaller lower bounds; 
without further assumptions $M_2> 45$ GeV. Using the relation \eqref{eq:mpconstr} $M_2$
is at least 100 GeV with or without mixing.
\begin{figure}[t]
\begin{center}
\includegraphics[scale=0.50]{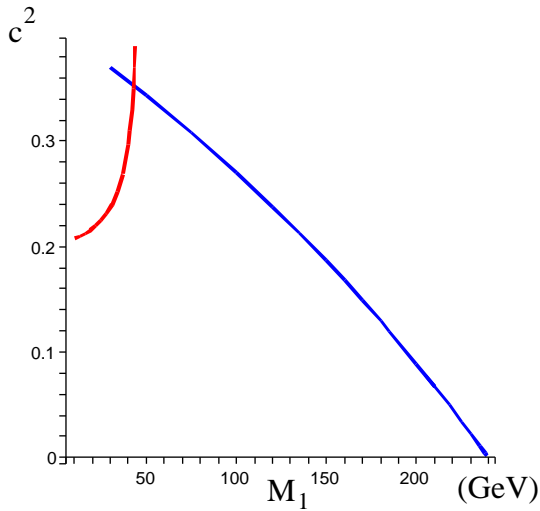}
\end{center}
\begin{center}
\caption{The maximum value of the cosine$^2$ of the mixing angle vs. the lighter neutral mass $M_1$. The right (blue curve) is derived from the gap equation and unitarity. The upper left (red) curve is from the width of the Z boson.}
\end{center}
\end{figure}
The mixing generates small, but non-vanishing coupling between the Z boson and the new 
lighter neutral fermion (e.g. the remnant of the singlet, it has $c^2$ part of a doublet.)
Therefore if it is light enough  it contributes to the invisible width of the Z boson
\begin{equation}
\Gamma(Z \rightarrow \bar{\Psi}_1 \Psi_1 )=\frac{\sqrt{2} {\mathrm G_F M_Z^3}}{6 \pi} 
  \left ( \frac{c^4}{4}\right ) \sqrt{1-\frac{4 M_1^2}{M_Z^2}} .
\end{equation}
The Z width is experimentally known at high precision and the pull factor is rather small
 \begin{equation}
\Gamma(Z  )=(2.4952 \pm 0.0023 ) \hbox{GeV} .
\end{equation}
We estimate the maximum possible room for new physics as 3$\sigma$ in the experimental Z width, $\Gamma_Z^{\hbox{new}}< $7MeV. In \cite{Pocsik} the minimum value of $\Gamma_Z^{\hbox{theory}}$ (at maximum $\sin^2 \theta_W$ and minimum $M_Z^2$ and $\alpha_S$)  was compared to the maximal experimental value, and gave a similar $3 \sigma$ window for new physics.
We see that $M_1$ masses well below $M_Z/2$ are still allowed for rather small mixing, see the (red) curve  on the left in Figure 2.

\section{Electroweak precision parameters}

The new fermions have direct interactions with the standard fermions \eqref{eq:Yukawa} and gauge bosons \eqref{eq:Llmix}.
The four-fermion couplings of the new particles to the light fermions
are weak; weaker than the corresponding ones in the standard model \cite{fcm}. The new couplings to the gauge bosons are the 
gauge couplings suppressed only by the ${\cal O} (1)$ mixing factors. Therefore the couplings to the light fermions which participate in the precision experiments, are suppressed compared to the couplings to the gauge bosons.
The new fermions thus mainly couple to the gauge boson self energies in the precision experiments.
In most of the solutions of the gap equation \cite{fcmgap} $M_+,M_2 \gg M_Z$ and expecting further $M_1 > M_Z$
we can give  a good estimate of the effects of new physics in terms of the general S, T and U parameters introduced by Peskin and Takeuchi \cite{stu}. 
We get a rough estimate of the loop effects if the mass of the
lighter neutral fermion is not far above the $Z$ mass.

The two relevant parameters, $S$ and $T$  defined  via the gauge boson self energies
\begin{eqnarray}
\alpha(M_Z) \, T & = & \frac{\Pi_{WW}^{{\rm new}}(0)}{M_W^2} -\frac{\Pi_{ZZ}^{{\rm new}}(0)}{M_Z^2}, \\
\frac{ \alpha(M_Z)}{4s_W^2 c_W^2 } \, S & = & \frac{ \Pi_{ZZ}^{{\rm new}}(M_Z^2)-\Pi_{ZZ}^{{\rm new}}(0)} {M_Z^2} 
-\frac{c_W^2-s_W^2}{c_W s_W} \frac{\Pi_{Z\gamma}^{{\rm new}}(M_Z^2)}{M_Z^2} -\frac{\Pi_{\gamma \gamma}^{{\rm new}}(M_Z^2)}{M_Z^2}
\label{eq:stuPT},
\end{eqnarray}
where $s_W^2=\sin^2{\theta}_W(M_Z)$ and $c_W^2=\cos^2{\theta}_W(M_Z)$ are  $\sin^2$ ($ \cos^2$) 
of the weak mixing angle. The $U$ parameter is suppressed by an extra factor of the weak gauge boson masses,
in most of the application $U \simeq 0$  and absent from newer parameterizations \cite{lep2}.
A more practical definition is based on the original $SU_L(2)$ and $U_Y(1)$
boson vacuum polarizations:
\begin{eqnarray}
\alpha(M_Z) T & = & \frac{1}{M_W^2} \left( \Pi_{33}^{{\rm new}}(0) -\Pi_{11}^{{\rm new}}(0) \right ), \\
\frac{ \alpha(M_Z)}{4s_W^2 c_W^2 } S & = & \Pi_{3Y}^{\prime \, {\rm new} }(0) .
\label{eq:stujo} 
\end{eqnarray}
The $\Pi$ functions are defined from the transverse gauge boson vacuum polarization amplitudes
expanded around zero   $ \Pi_{ab}(q^{2})\simeq\Pi_{ab}(0)+q^{2}\Pi'_{ab}(0)+1/2 \cdot q^{2} \Pi''_{ab}(0)+...$, (a,b = 1,3,Y).

The experimental data determines $S$ and $T$  (without fixing $U=0$)
\begin{eqnarray}
S & = & -0.10 \pm 0.10 \; (-0.08 ), \label{Sexp} \\
T & = & -0.08 \pm 0.11 \; (+0.09),   \label{Texp}
\label{eq:stuexp}
\end{eqnarray}
where the central value assumes $M_H=117 $ GeV and in parentheses the difference is shown for $M_H=300 $ GeV.
In our model the Higgs mass of the fit is understood as the contribution of a composite Higgs particle with the given mass.

The contributions of the new sector to the gauge boson vacuum polarizations are 
fermion loops with generally two non-degenerate masses $m_{a}$ and $m_{b}$.
In the low energy effective model we have preformed the calculation
with  a 4-dimensional momentum cutoff $\Lambda$.
The coupling constants are defined in the usual manner $L^{I}\sim V_{\mu}\bar{\Psi}\left(g_{V}\gamma^{\mu}+g_{A}\gamma_{5}\gamma^{\mu}\right)\Psi$
\begin{equation}
\Pi(q^{2})=\frac{1}{4\pi^{2}}\left( g_{V}^{2} \,\tilde{\Pi}_{V}+g_{A}^{2} \, \tilde{\Pi}_{A}\right)\label{eq:pivv} .
\end{equation}


The electroweak parameters depend on the values and derivatives of
the $\Pi$ functions at $q^{2}=0$
\begin{eqnarray}
\tilde{\Pi}_{V}(0) & = & \frac{1}{4} (m_{a}^{2}+m_{b}^{2} )
-\frac{1}{2}\left(m_{a}-m_{b}\right)^{2}
\ln\left(\frac{\Lambda^{2}}{m_{a}m_{b}}\right)- 
\label{eq:piv0} \\
& & -\frac{ m_{a}^{4}+m_{b}^{4} -2m_a m_b \left(m_{a}^{2}+m_{b}^{2}\right)}{4\left(m_{a}^{2}-m_{b}^{2}\right)}
\ln\left(\frac{m_{b}^{2}}{m_{a}^{2}}\right).
\label{eq:piv+a0}  \nonumber
\end{eqnarray}
The first derivative is
\begin{eqnarray}
\tilde{\Pi}'_{V}(0) & \!\!=\!\!\! & -\frac{2}{9}-
\frac{4 m_{a}^{2}m_{b}^{2} -3 m_a m_b \left(m_{a}^{2}+m_{b}^{2}\right) }{6\left(m_{a}^{2}-m_{b}^{2}\right)^{2}}
+\frac{1}{3} \ln\left(\frac{\Lambda^{2}}{m_{a}m_{b}}\right) + 
\label{eq:piv+av0} \\ & & +\frac{\left(m_{a}^{2}+m_{b}^{2}\right)\left(m_{a}^{4}-4m_{a}^{2}m_{b}^{2}+m_{b}^{4}  \right) +6m_a^3 m_b^3}{6\left(m_{a}^{2}-m_{b}^{2}\right)^{3}}
\ln\left(\frac{m_{b}^{2}}{m_{a}^{2}}\right).
 \nonumber
\end{eqnarray}
For completeness we give the second derivative, too. It can be used to calculate further precision parameters e.g. extra two parameters introduced by Barbieri et al. \cite{lep2,stu6}.
\begin{eqnarray}
\tilde{\Pi}''_{V}(0) & = & \frac{\left(m_{a}^{2}+m_{b}^{2}\right)\left(m_{a}^{4}-8m_{a}^{2}m_{b}^{2}+m_{b}^{4}\right)}{8\left(m_{a}^{2}-m_{b}^{2}\right)^{4}}+
\frac{ m_{a}m_{b} \left(m_{a}^{4}+10m_{a}^{2}m_{b}^{2}+m_{b}^{4}\right)}{6\left(m_{a}^{2}-m_{b}^{2}\right)^{4}} -\\
& & -\frac{ m_a^3 m_b^3 \left(  3m_a m_b-2m_a^2 -2 m_b^2 \right) }{2 \left(m_{a}^{2}-m_{b}^{2}\right)^{5}}\ln\left(\frac{m_{b}^{2}}{m_{a}^{2}}\right) .
\label{eq:piv+avv0}
\end{eqnarray}
We get the functions for axial vector coupling by flipping exactly one of the masses in the previous results
($ m_a\rightarrow m_a$ and $m_b \rightarrow -m_b$).
The method of our calculation has nice properties: it has no quadratic divergence as expected;
it fulfills gauge invariance in two aspects, $\Pi_V(m_a,m_a,0)=0$ and the complete $\Pi$ function is transverse,
the coefficients of the $g_{\mu \nu}$ and $-p_\mu p_\nu /p^2$ parts are equal.

The values of the vacuum polarizations for identical masses $(m_{b}=m_{a})$
are  smooth limits and agree with direct calculation.
\begin{eqnarray}
\tilde{\Pi}_{V\!}(0)=0, \quad 
\tilde{\Pi}'_{V\!}(0)=-\frac{1}{3}+\frac{1}{3} \ln\left(\frac{\Lambda^{2}}{m_{a}^2}\right),\quad  
\tilde{\Pi}''_{V\!}(0)=\frac{2}{15} \frac{1}{m_a^2}.
\end{eqnarray}
The ${S}$ parameter is then given by (for the sake of simplicity the index $V$ is omitted)
\begin{equation}
{S}=\frac{1}{\pi}\left(+\tilde{\Pi}'(M_{+},M_{+},0) -c^{4}\tilde{\Pi}'(M_{1},M_{1},0) -s^{4}\tilde{\Pi}'(M_{2},M_{2},0) -2s^{2}c^{2}\tilde{\Pi}'(M_{2},M_{1},0)\right).
\label{eq:spar}
\end{equation}
The first three terms cancel the divergent contribution of the last one.

The ${T}$ parameter  related to $\Delta\rho$ is 
\begin{eqnarray}
{T} & = & \frac{1}{4 \pi s_W^2 M_{W}^{2}}\left[+\tilde{\Pi}(M_{+},M_{+},0) +c^{4}\tilde{\Pi}(M_{1},M_{1},0) +s^{4}\tilde{\Pi}(M_{2},M_{2},0)+\right.\nonumber \\
 &  &  \left. 
+2s^{2}c^{2}\tilde{\Pi}(M_{2},M_{1},0) -2c^{2}\tilde{\Pi}(M_{+},M_{1},0) -2s^{2}\tilde{\Pi}(M_{+},M_{2},0)\right]
.\label{eq:tpar}
\end{eqnarray}

\section{Numerical results}

There are 3 free parameter in the model to confront with experiment.
These can be chosen the three dimensionful four-fermion couplings $\lambda_{1,\,2,\,3}$, or more practically the two physical neutral masses $M_1$, $M_2$ and the mixing angle,   $c^{2}=\cos^{2}\phi$. For the cutoff $\Lambda \simeq 3$ TeV there is a maximum value for the masses, $M_1\leq 240$ GeV and for $c^2$ as a function of $M_1$, see Figure 1.
The mass of the charged fermion is given by the solution of the gap equations, the value of $M_+$ is close to, but not equal to  $c^{2}M_{1}+s^{2}M_{2}$.

If there is no real mixing $c^{2}=0$; or if $M_{1}=M_{2}=M_{+}$,
then there is one degenerate vector-like fermion doublet and a decoupled
singlet, and ${S}$ and ${T}$ vanish explicitely. In this case
the new sector does not violate $SU_{L}(2)$ and there is an exact
custodial symmetry. Increasing the mass difference in the remnants
of the original doublet by increasing the $\left|M_{1}-M_{2}\right|$
mass difference and/or moving away from the non-mixing case $c^{2}=0,$
results in increasing ${S}$ and ${T}$. For small violation
of the symmetries ${S}$ and ${T}$ are expected to be small.
In case of relatively small masses the oblique parameters are understood
as rough estimates, but still in agreement with experiment.

\begin{figure}[t]
\begin{center}
\includegraphics[scale=0.50]{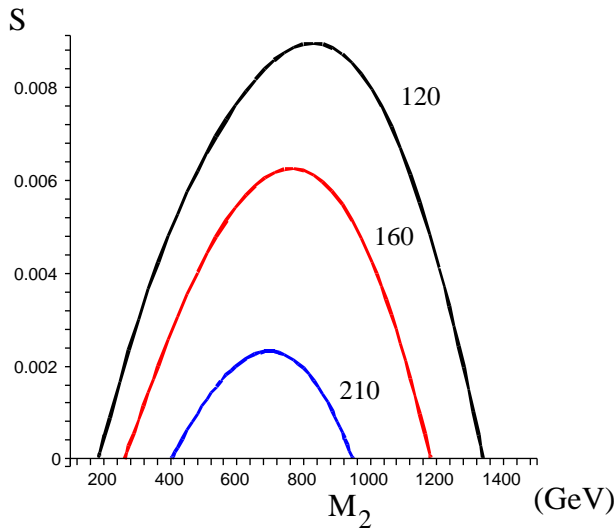}
\caption{The maximum value of the ${S}$ parameter vs. $M_{2}$ for 
$M_{1}=120,\, 160,\, 210$  GeV. The 95 \% C.L. bounds [-0.296, 0.096] are
outside the figure.}
\end{center}
\end{figure}

Generally the $S$ parameter depends only on the masses
of the new particles and the mixing angle.
For the solutions of the gap equations fulfilling perturbative unitarity 
the $ S $ parameter is always positive
and far below the 95 \% C.L. For a given $M_1, \, M_2$  $ S $ increases with increasing $c^2$
and maximal for the highest $c^2$. This maximum value of the $S$ parameter is plotted against
$M_2$ for three given $M_1$ in Figure 2. The small value of $S$ does not constrain the parameters of the model.

The value of the $T$ parameter is always positive.
The ${T}$ parameter (\ref{eq:tpar})   sensitive
to the differences and ratios of the masses $M_{1,\,2,\,+}$.
$T$ still varies for a given  $(M_1, M_2) $ pair
depending on $M_+$ or equally on  $c^2$;
$T$ is maximal for largest mass difference, for the largest $c^2$ allowed by the gap equations and perturbative unitarity. 
The $T$ parameter can always be in agreement with experiment for any  $(M_1, M_2) $ pair for small mixing, for $c^2=0$ the $T$ parameter vanishes identically.
We plotted the worst case in the $(M_1,M_2)$ plane, the possible maximum value of the $T$ parameter; it is given by the maximum $M_2-M_+$ mass difference or equally for maximal $c^2$.

\begin{figure}
\begin{center}
\includegraphics[scale=0.50]{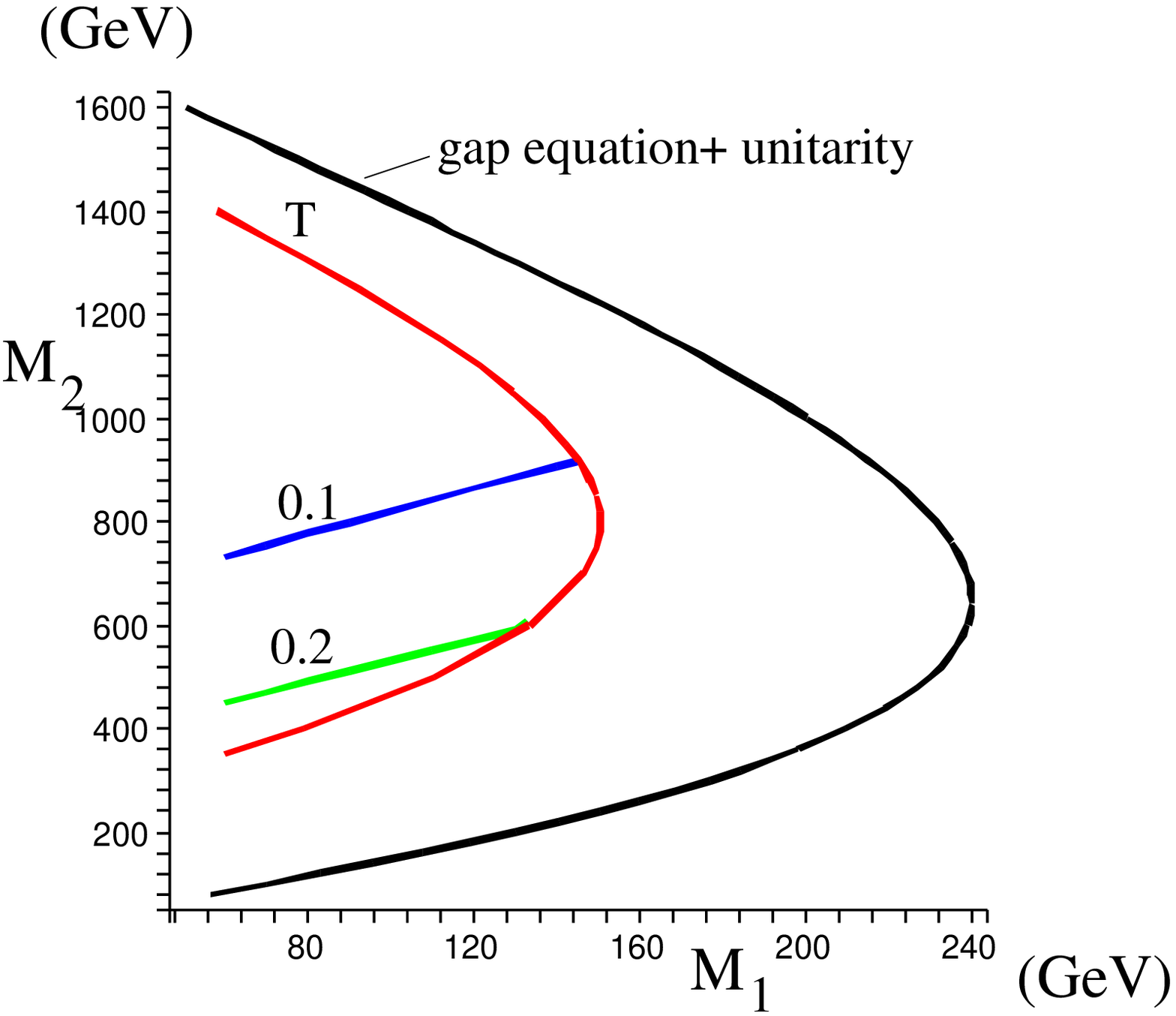}
\includegraphics[scale=0.50]{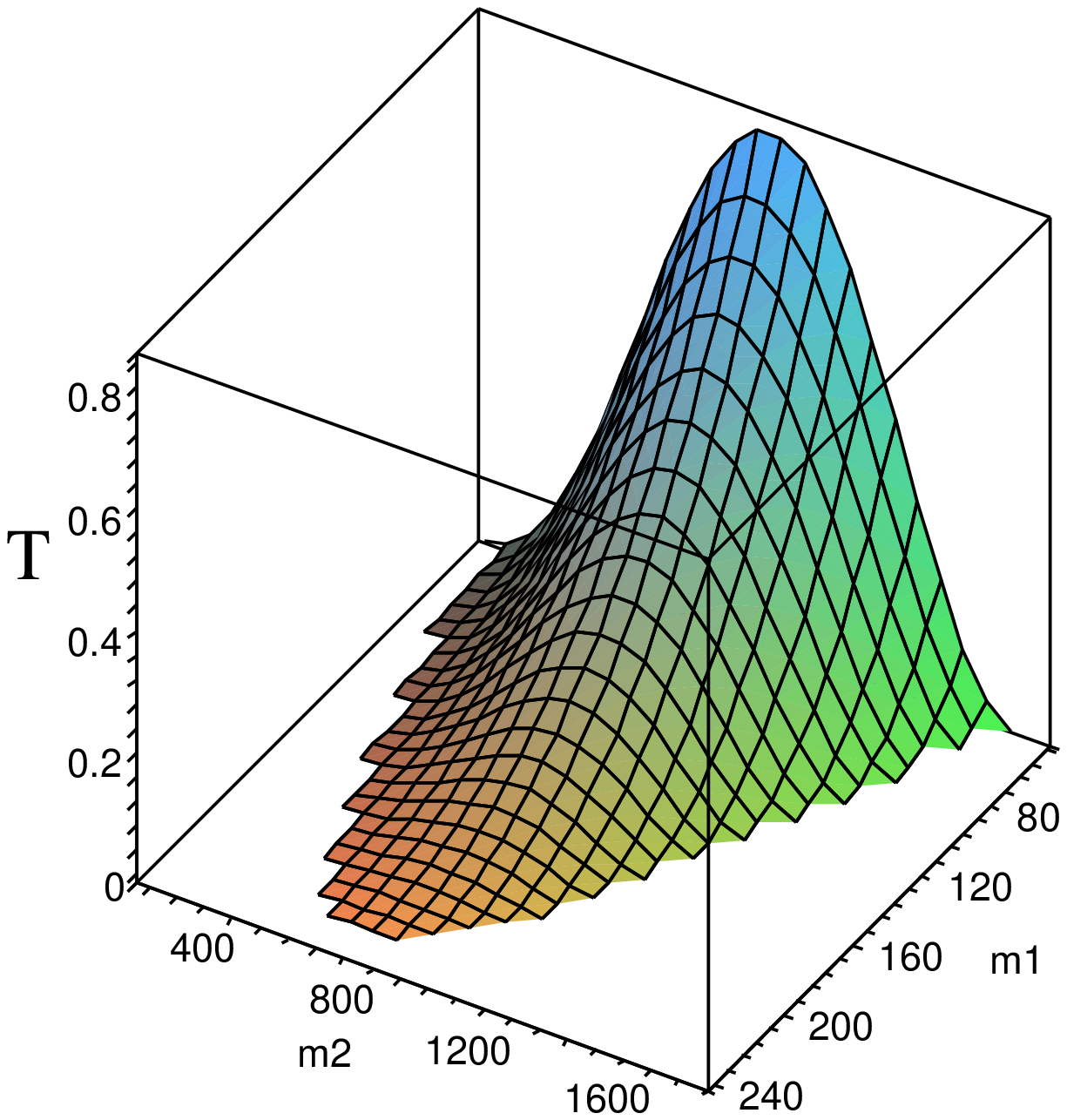}
\end{center}

\noindent \caption{ Constraints on the  $(M_1,M_2)$ plane. 
The solution of the gap equations respecting perturbative unitarity are inside the outer curve. The inner curve shows the region, where the ${T}$ parameter  gives the maximum value of $c^2$ at 95 \% C.L.. Below the 0.1 (blue) and 0.2 (green) line $c^2$ can exceed 0.1 and 0.2. The right panel shows the maximum value of $T$ vs. $(M_1,M_2)$. }
\end{figure}

If the Higgs is heavy, e.g. $M_{H}=300$ GeV (\ref{Sexp}, \ref{Texp}) the central
value of ${S}$ decreases and  ${T}$ increases compared to the light Higgs
case. The  ${S}$  parameter still in agreement with 
the predictions of the model.
Incrasing the Higss mass the Standard Model moves away in the (S,T) plane 
from the experimentally allowed ellipse, see \cite{LEPEWWG}.
The negative contribution ($-.09 $) of the heavy Higgs to the $ {T} $  parameter can be compensated by  the positive $ {T} $ contribution of the new fermions with considerable 
mass difference. For example (160, 800) GeV and the largest  mixing $ c^2\sim 0.115 $ allowed by the gap equations and unitarity gives $\Delta T\simeq 0.1$.
Even heavier Higgs boson can be compensated as can be read off from Figure 4.
Non-degenerate vector-like fermions with reasonable mixing allow a space for heavy Higgs 
in the precision tests of the Standard Model.

\section{Conclusions}

We have calculated the oblique corrections in an extension of the
Standard Model based on vector-like weak singlet and doublet fermions.
Due to non-diagonal condensate (\ref{eq:mixed cond}) symmetry breaking
mixing occurs between the singlet and the neutral component of the
doublet. The oblique corrections were presented in the Peskin-Takeuchi formalism 
\cite{stu}. 
The corrections depend on the masses  of the new fermions ($M_{1,2}$) and the mixing angle.
The ${S}$  parameter is always in agreement with experiment at 95 \% C.L. even for heavier
Higgs mass.
The $T$ parameter measures the custodial symmetry breaking, the custodial symmetry is
exact in the new sector if there is no physical mixing: $c^{2}=0$
or $M_{1}=M_{2}$. The gap equations and perturbative unitarity already constrains the mass range and the mixing of the model for a given cutoff.
For $\Lambda=3 $ TeV the lighter neutral mass must be  smaller than approximately 240 GeV and
the cosine of the mixing angle is bounded above.
The $T$ parameter further constrains $c^2$ for relatively small masses ($M_1<150 $ GeV), but
there is always a small enough $c^2$, which produces small $T$ parameter.
This modification of of the standard model nicely accommodates a composite heavy Higgs in the
precision electroweak test of the standard model. The lightest new fermion
is stable and a good dark matter candidate. The model can be tested
at LHC in the Drell-Yan process \cite{fcm} or via jetmass analysis
\cite{jetmass}. 


\appendix
\section{Regularization with momentum cutoff}

There are low energy theories, like the Fermion Condensate Model, which have an intrinsic cutoff, i.e. the upper bound of the model.
The naive calculation of divergent Feynman graphs with a momentum cutoff is thought to break continuous symmetries of the model. In this case the gauge invariance of the two point function with two different fermion masses in the loops can be reconstructed by subtractions leading to finite ambiguity. To avoid these problems we used dimensional regularization in $d=4-2\epsilon$ and identified the poles at $d=2$ with quadratic divergencies while  the poles at $d=4$ with logarithmic divergencies \cite{hagiwara}. Carefully calculating the one and two point Passarino-Veltman functions in the two schemes the divergencies are the following in the momentum cutoff regularization
\begin{eqnarray}
4\pi \mu^2 \left(\frac{1}{\epsilon -1} +1 \right ) &=& \Lambda^2,  \\
\frac{1}{\epsilon}-\gamma_E+ \ln \left ( 4\pi \mu^2 \right )+1 &=& \ln \Lambda^2,
\end{eqnarray}
where $\mu$ is the massscale of dimensional regularization. The finite part of a divergent quantity is defined by
\begin{equation}
f_{\rm finite}= \lim_{\epsilon \rightarrow 0} 
\left [ f(\epsilon)-R(1) \left(\frac{1}{\epsilon -1} +1 \right ) -R(0) \left (\frac{1}{\epsilon}-\gamma_E+ \ln  4\pi +1 \right )   \right ],
\end{equation}
where $R(1)$, \, $R(0)$ are the residues of the poles at $\epsilon=1,\, 0$ respectively.

We have found that contrary to the expectations  the ambiguity of the cutoff regularization scheme is coming from the replacement of $l_\mu l_\nu \rightarrow g_{\mu \nu} l^2 /4$  and not from shifting the loop-momentum $(l)$ \cite{new}.

\bibliography{EFHPP}
\bibliographystyle{JHEP}

\providecommand{\href}[2]{#2}\begingroup\raggedright\endgroup

\end{document}